\documentclass[aps,twocolumn,superscriptaddress,showpacs]{revtex4}%

\usepackage{amssymb}
\usepackage{amsmath}
\usepackage{graphicx}
\usepackage[normalem]{ulem}
\usepackage[dvips]{color}
\usepackage[normalem]{ulem} 
\usepackage[dvips]{color} 
\renewcommand\sout{\bgroup \color{red} \ULdepth=-.5ex \ULset}

\def\es0{$E_{sym}(\rho_0)$~}
\begin{document}

\title{Probing the neutron-skin thickness by photon production from reactions induced by intermediate-energy protons}
\author{Gao-Feng Wei}\email[Email address: ]{wei.gaofeng@foxmail.com}
\affiliation{Department of Applied Physics, Xi'an JiaoTong University, Xi'an 710049, China}
\affiliation{School of Physics and Mechatronics Engineering, Xi'an University of Arts and Science, Xi'an, 710065, China}
\affiliation{Department of Physics and Astronomy, Texas A$\&$M University-Commerce, Commerce, TX 75429-3011, USA}


\begin{abstract}
Photon from neutron-proton bremsstrahlung in p+Pb reactions is examined as a potential
probe of the neutron-skin thickness in different centralities and at different proton
incident energies. It is shown that the best choice of reaction environment is about
140MeV for the incident proton and the 95\%-100\% centrality for the reaction system
since the incident proton mainly interacts with neutrons inside the skin of the target
and thus leads to different photon production to maximal extent. Moreover, considering
two main uncertainties from both photon production probability and nucleon-nucleon
cross section in the reaction, I propose to use the ratio of photon production
from two reactions to measure the neutron-skin thickness because of its cancellation
effects on these uncertainties simultaneously, but the preserved about 13\%-15\%
sensitivities on the varied neutron-skin thickness from 0.1 to 0.3fm within the
current experimental uncertainty range of the neutron-skin size in $^{208}$Pb.

\end{abstract}


\maketitle

The neutron-skin of nuclei is a fundamental physical quantity in nuclear physics, and has
received considerable attention due to its importance in determining the structure of
neutron-rich nuclei in nuclear physics and the property of neutron-rich matter in
astrophysics. To determine the neutron-skin thickness of nuclei, one should known the proton
density distribution and neutron density distribution, and then determines the corresponding
neutron-skin thickness by calculating the root-mean-square (rms) radius difference between
proton and neutron. Presently, the proton rms radius can be determined precisely, typically
with an error of 0.02fm or better for many nuclei \cite {Fricke95,Brown00,Sun10}; the neutron
rms radius is much less well-known although many efforts have been devoted to probing the neutron
density distribution by theoretical and experimental methods such as the nucleon elastic
scattering \cite{Hoff80,Kar02,Clark03,Zen10}, the inelastic excitation of the giant dipole
and spin-dipole resonances \cite{Kra99,Kra04}, the pygmy dipole resonance \cite {Kli07,Car10}
and experiments in exotic atoms \cite{Trz01,Fri03,Jas04,Fri05,Klos07,Fri09}. This is because
almost all of these probes are hadronic ones and need model assumptions to deal with the strong
force introducing possible systematic uncertainties even if some of them reach small errors
\cite {Roca11}. In this situation, the Parity Radius Experiment (PREX-I) at the Jefferson
Laboratory (J-Lab)\cite{Prex1} has been performed to measure the neutron-skin thickness of
$^{208}$Pb using parity violating $e$-Pb scattering, the measured value of $0.33^{+0.16}_{-0.18}$
fm in $^{208}$Pb obviously differs from previous value of $0.11\pm 0.06$ fm of $^{208}$Pb from
$\pi^+$-Pb scattering \cite{Fri12} albeit largely overlapping with each other within error bars.
However, the obtained results from PREX-I experiment suffer from large uncertainties although the
PREX-I experiment aims to a model-independent measurement of the neutron-skin thickness of
$^{208}$Pb. It is interesting, however, to note that the neutron-skin for $^{208}$Pb as thick as
$0.33+0.16$ fm reported by the PREX-I experiment can not be ruled out within a relativistic
mean-field model \cite{Fatt13}. This situation stimulated the J-Lab to plan to remeasure the
neutron-skin thickness of $^{208}$Pb and $^{48}$Ca, i.e., the PREX-II experiment and the Calcium
Radius Experiment, which are expected to provide more accurate neutron-skin thickness for
$^{208}$Pb and $^{48}$Ca \cite{Hor14}. While waiting the experimental data, theoretical efforts
on this problem are required to indicate what are the sensitive probes of the neutron-skin
thickness especially those of non-hadronic ones.

\begin{figure}[h]
\centerline{\includegraphics[scale=0.35]{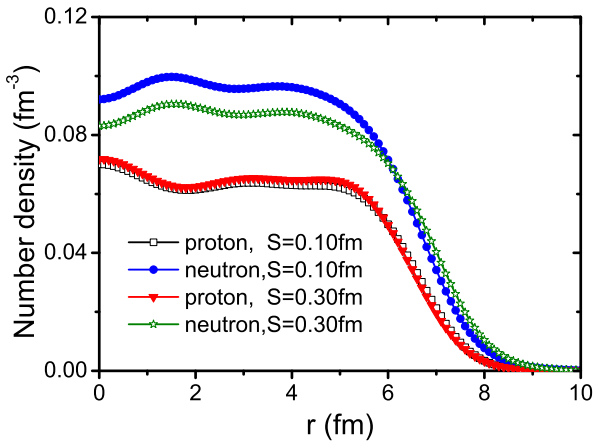}} \caption{(Color
online) The neutron and proton density profiles for $^{208}$Pb target with
the neutron-skin thickness of 0.1 and 0.3 fm. } \label{density}
\end{figure}

Similar to electrons, photons interact with nucleons only electromagnetically, and they
escape almost freely from the nuclear environment once produced. In fact, photon production
in heavy-ion reactions has been extensively studied in experiment and theory \cite {Bert88,
Nif90,Cass90}. For example, the hard photon from neutron-proton bremsstrahlung is employed
to probe the nuclear caloric curve \cite{Orte04}, the dynamics of nucleon-nucleon interactions
\cite{Schu97,Mart99,Ente02}, the time-evolution of the reaction process before nuclear
break-up \cite{Orte06} as well as the space-time
extent of the photon emitting sources \cite{TAPS94}; and the soft photon from giant dipole
resonances in heavy-ion reactions is used to study the symmetry potential term of the
nucleon-nucleon interactions \cite{Giu06}. A natural question is whether the photon can be
used as a potential sensitive probe of the neutron-skin thickness in nuclear reactions. Before
answer this question, let's first initialize the $^{208}$Pb target with different density
distribution corresponding to two different neutron-skin size of $S$=0.10 and 0.30fm within
the current experimental uncertainty range of the neutron-skin size of $^{208}$Pb, which are
predicted by Hartree-Fock calculations based on the MSL model \cite{Che09,Che10}. Different
values of neutron-skin thickness can be obtained by changing only the value of $L$ in the MSL0
force \cite{Che10} while keeping all the other macroscopic quantities the same. Shown in
Fig. \ref{density} are the density profiles corresponding to the neutron-skin thickness of
0.1 and 0.3 fm for $^{208}$Pb target \cite{Wei14,Wei15}, the proton distributions are almost
identical, while the neutrons distribute differently in the two cases considered.

To answer the question mentioned above, one has to confront two main uncertainty
factors because they can significantly influence the photon production in our
reaction model IBUU \cite {IBUU}. One is the in-medium nucleon-nucleon cross
section defined as,
\begin{equation}\label{xsection}
\sigma^{\rm NN}_{\rm med}=\sigma^{\rm NN}_{\rm free}\big(\frac{\mu^{*}_{\rm NN}}{\mu_{\rm NN}}\big)^{2},
\end{equation}
where the $\mu^{*}_{\rm NN}$ and $\mu_{\rm NN}$ are the in-medium and free-space reduced
nucleon-nucleon mass. The scaling factor $(\frac{\mu^{*}_{\rm NN}}{\mu_{\rm NN}})^{2}$
reduces significantly the relative cross sections of nucleon-nucleon collisions
due to the momentum dependence of the nuclear interactions \cite{Bao05}. Another is the photon
production probability, since this probability is very small, i.e., just one photon
producing roughly in a thousand nucleon-nucleon collisions. Therefore, photon
production in dynamical calculations of nuclear reactions at intermediate energy
is usually treated in a perturbative manner \cite{Bert88,Cass90}. In this approach,
one calculates the photon production as a probability at each proton-neutron collision
and then sums over all such collisions over the entire history of the reaction
\cite{Yong08,Yong11}. Two kinds of probability formula are commonly used to predict
the photon production in nuclear reactions. One is based on the semiclassical hard
sphere collision model \cite{Bert88,Cass90}, its definition is,
\begin{equation}\label{pa}
p^{a}_{\gamma}\equiv \frac{dN}{d\varepsilon_{\gamma}}=1.55\times 10^{-3}\times
\frac{1}{\varepsilon_{\gamma}}(\beta^{2}_{i}+\beta^{2}_{f}),
\end{equation}
where $\varepsilon_{\gamma}$ is the energy of emitting photon, $\beta_{i}$ and $\beta_{f}$ are
the initial and final velocities of the proton in the proton-neutron center of mass frame.
Another is based on the one-boson exchange model involving more quantum mechanical effects
\cite{Gan94} as follows,
\begin{equation}\label{pb}
p^{b}_{\gamma}\equiv \frac{dN}{d\varepsilon_{\gamma}}=2.1\times 10^{-6}\times
\frac{(1-y^{2})^{\alpha}}{y},
\end{equation}
where $y=\varepsilon_{\gamma}/E_{max}$, $\alpha$=0.7319-0.5898$\beta_{i}$, and $E_{max}$ is the
energy available in the center of mass of the colliding proton-neutron pairs.
\begin{figure}[h]
\centerline{\includegraphics[scale=0.35]{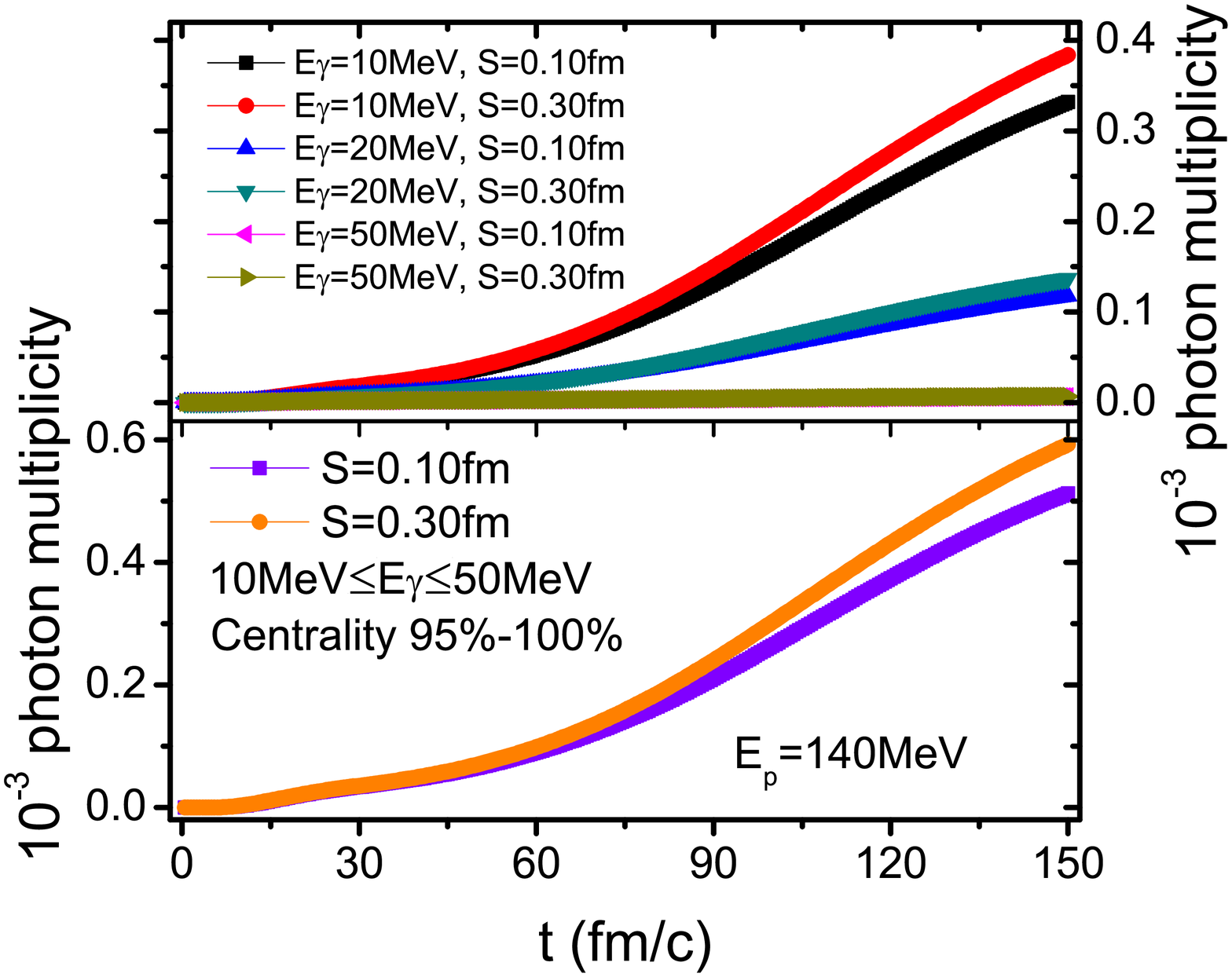}} \caption{(Color
online) The time evolution of photon multiplicity with different energies (upper panel) and
total photon multiplicity (lower panel) in p+Pb reaction with 95\%-100\% centrality at the
proton incident energy of 140MeV within the neutron-skin thickness of 0.1 and 0.3fm. The
probability formula $p^{b}_{\gamma}$ and in-medium nucleon-nucleon cross section are used.}
\label{multi}
\end{figure}
\begin{figure}[h]
\centerline{\includegraphics[scale=0.35]{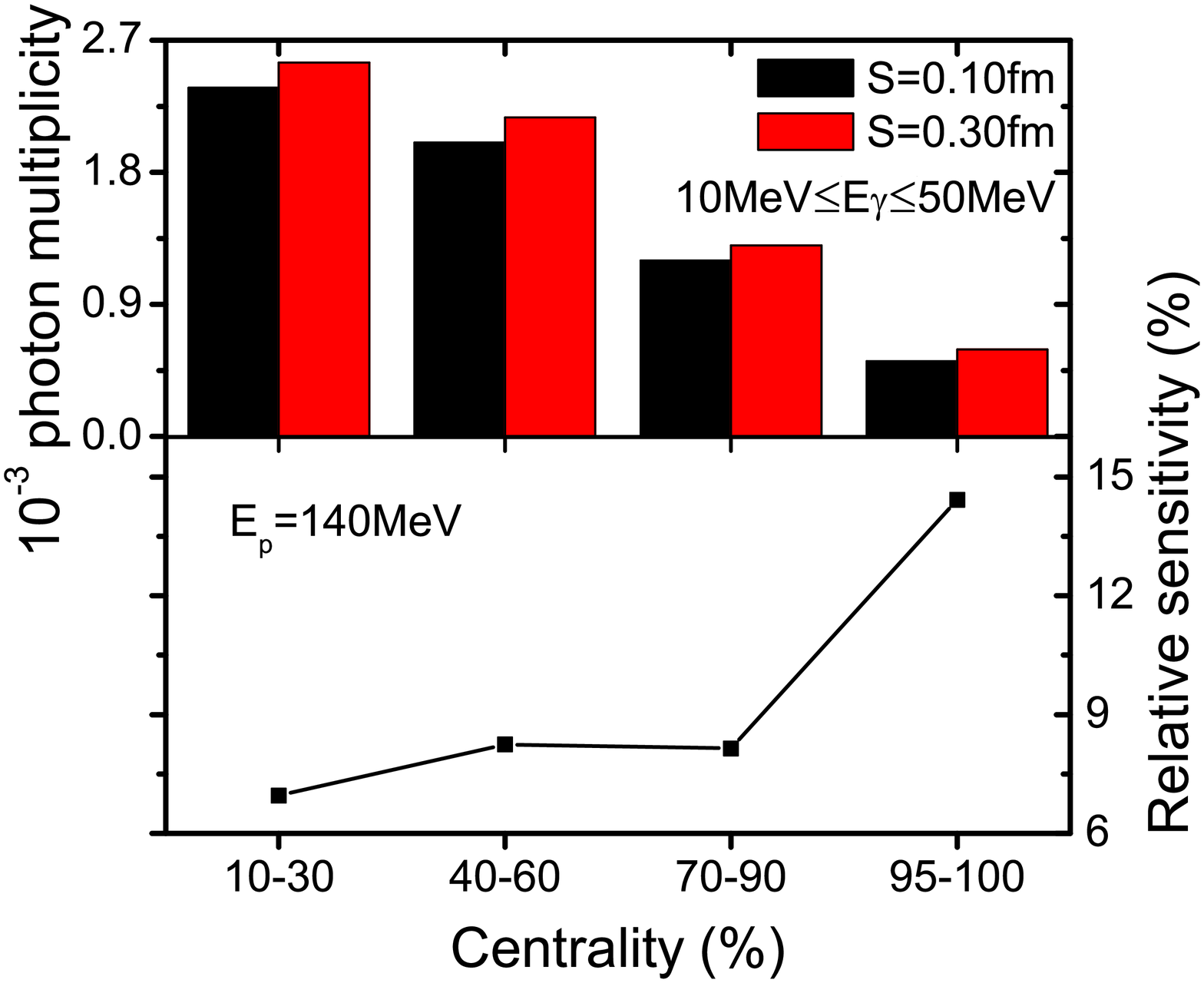}} \caption{(Color
online) Total photon multiplicity and the corresponding relative sensitivity on the neutron-skin
thickness in different centralities but the given proton incident energy of 140MeV within the neutron-skin
thickness of 0.10 and 0.30fm. The probability formula $p^{b}_{\gamma}$ and in-medium nucleon-nucleon
cross section are used.}\label{cent}
\end{figure}
\begin{figure}[h]
\centerline{\includegraphics[scale=0.35]{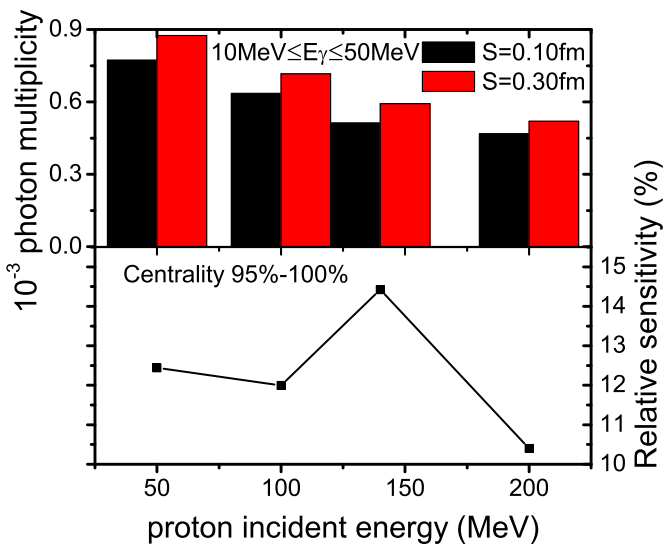}} \caption{(Color
online) Total photon multiplicity and the corresponding relative sensitivity on the neutron-skin
thickness in different proton incident energies but the given centrality of 95\%-100\% within the neutron-skin
thickness of 0.10 and 0.30fm. The probability formula $p^{b}_{\gamma}$ and in-medium nucleon-nucleon
cross section are used.} \label{energy}
\end{figure}

Now let's check the sensitivities of photon production from neutron-proton bremsstrahlung
on the neutron-skin thickness in p+Pb reaction. Shown in Fig. \ref{multi} is the time evolution
of photon multiplicity with different energies (upper panel) and total photon multiplicity
(lower panel) in p+Pb reaction with 95\%-100\% centrality and at the proton incident energy of
140MeV within the neutron-skin thickness of 0.1 and 0.3fm. Here, the centrality is defined as
the percent of impact parameter over the size of reaction system. First, it can be seen from the
upper panel of Fig. \ref{multi} that the photon multiplicity is decreasing with the photon energy
increasing, and thus the production of photon with energy beyond about 50MeV can be ignored in
the intermediate energy p+Pb reaction. Second, the photon multiplicity with the thicker
neutron-skin is larger than that with the thinner neutron-skin especially for those of lower
energy photon, this is because the larger neutron densities inside the thicker neutron-skin get
these neutrons with higher probability to repeatedly collide with incident proton and thus leads
to higher photon production, especially for emitting lower energy photon. However, considering that
photon production is insufficient large after all, I thus check the sensitivity of total photon
multiplicity with energy from about 10 to 50MeV on the neutron-skin thickness. It can be seen from
the lower panel of Fig. \ref{multi} that the total photon multiplicity is also sensitivity to the
neutron-skin thickness, and shows about 15\% relative sensitivity. Nevertheless, is the 140MeV
the best proton incident energy in probing the neutron-skin thickness using photon production
in p+Pb reaction, and whether the 95\%-100\% centrality is the best choice for the reaction
system? Shown in Figs. \ref{cent} and \ref{energy} are the total photon multiplicity and
corresponding relative sensitivity on the neutron-skin thickness in different centralities but
the given proton incident energy of 140MeV, and in different proton incident energies but the
given centrality of 95\%-100\%, respectively, within the neutron-skin thickness of 0.10 and 0.30fm.
It can be seen that the best choice of reaction environment is about 140MeV for the incident proton and
the 95\%-100\% centrality for the reaction system since the incident proton mainly interacts with
the neutron inside the skin of the target and thus leads to different photon production to maximal
extent. Certainly, with the incident proton energy increasing the higher photon production may be
reachable, but the produced $\pi^{0}$ mesons can also produce photon and thus bring in more
complicated physics process. Therefore, I employ the proton incident energy of 140MeV and the
centrality of 95\%-100\% as the best reaction environment in probing the neutron-skin thickness
using the photon production.

\begin{figure}[h]
\centerline{\includegraphics[scale=0.35]{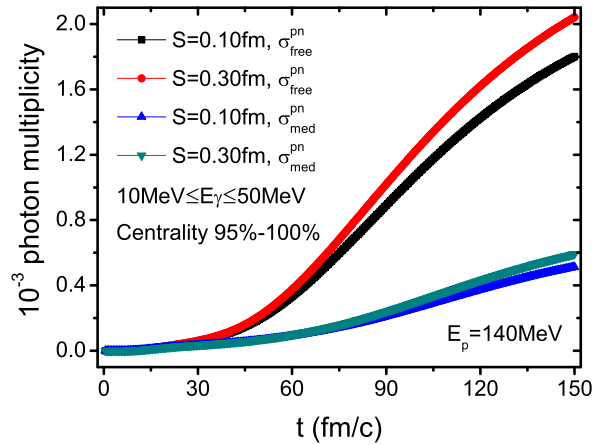}} \caption{(Color
online) The time evolution of total photon multiplicity with free-space and in-medium
nucleon-nucleon cross section in the proton incident energy of 140MeV and the centrality
of 95\%-100\% within the probability formula $p^{b}_{\gamma}$ and
neutron-skin thickness of 0.10 and 0.30fm.} \label{med}
\end{figure}
\begin{figure}[h]
\centerline{\includegraphics[scale=0.35]{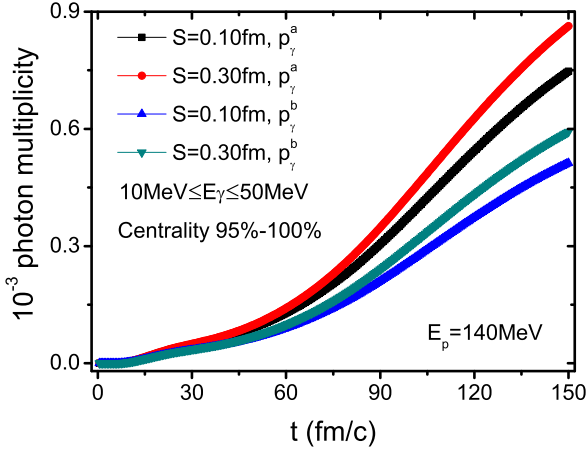}} \caption{(Color
online) The time evolution of total photon multiplicity with probability formulae
$p^{a}_{\gamma}$ and $p^{b}_{\gamma}$ in the proton incident energy of 140MeV and the
centrality of 95\%-100\% within the in-medium nucleon-nucleon cross section and
neutron-skin thickness of 0.10 and 0.30fm.} \label{prob}
\end{figure}
\begin{figure}[h]
\centerline{\includegraphics[scale=0.35]{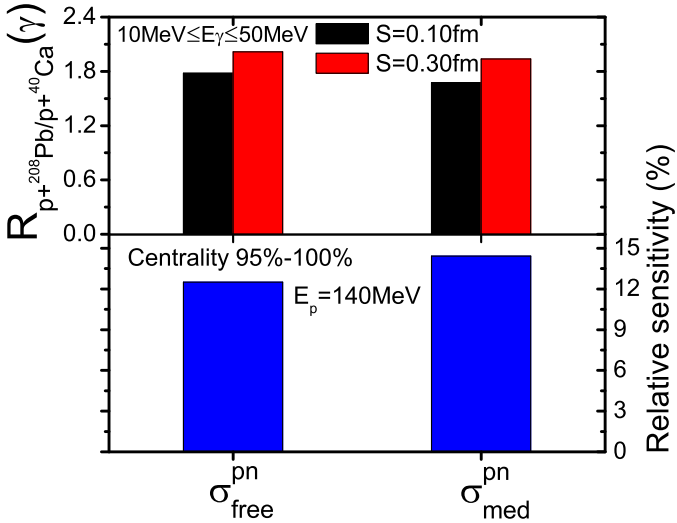}} \caption{(Color
online) The ratio of photon multiplicity from two reactions and the corresponding
relative sensitivity on the neutron-skin thickness with free-space and in-medium nucleon-nucleon
cross section within the probability formula $p^{b}_{\gamma}$ and neutron-skin thickness
of 0.10 and 0.30fm.} \label{med2}
\end{figure}
\begin{figure}[h]
\centerline{\includegraphics[scale=0.35]{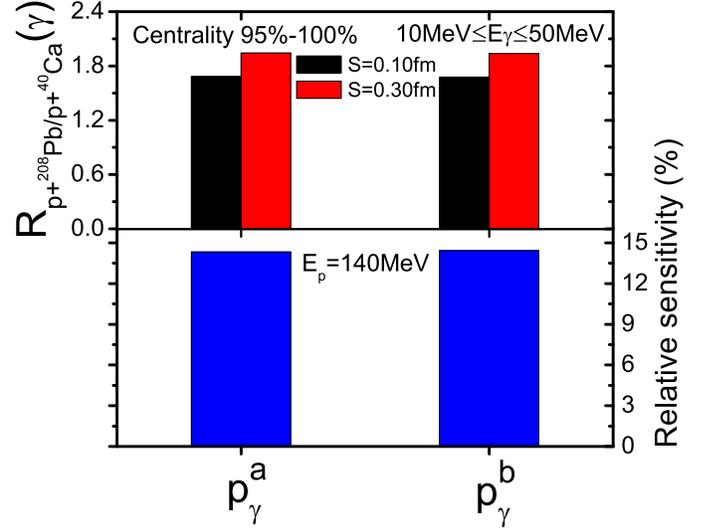}} \caption{(Color
online) The ratio of photon multiplicity from two reactions and the corresponding
relative sensitivity on the neutron-skin thickness with different photon production probability
within the in-medium nucleon-nucleon cross section and neutron-skin thickness of 0.10
and 0.30fm.} \label{prob2}
\end{figure}
However, the influence of two main uncertainties from both nucleon-nucleon
cross section and photon production probability may change the effects of the
neutron-skin thickness on the total photon multiplicity. Shown in Figs. \ref{med}
and \ref{prob} are the time evolution of total photon multiplicity with different
nucleon-nucleon cross section, and with different photon production probability,
respectively, within the neutron-skin thickness of 0.10 and 0.30fm. First, the
total photon multiplicity with free-space nucleon-nucleon cross section is higher
than that with in-medium nucleon-nucleon cross section because the scaling factor
$(\frac{\mu^{*}_{\rm NN}}{\mu_{\rm NN}})^{2}$ reduces significantly the relative cross sections
of nucleon-nucleon collisions \cite{Bao05}. Second, the total photon multiplicity with
probability formula $p^{a}_{\gamma}$ is higher than that with probability formula
$p^{b}_{\gamma}$ similar to the results reported in previous Refs. \cite{Yong08,Yong11,Gan94}.
It is fortunate to see that the sensitivity of total photon multiplicity on the neutron-skin
thickness is not changed no matter how the nucleon-nucleon cross section and/or photon
production probability change. However, the influence of these uncertainties on photon
production is much more than the effects from the neutron-skin thickness. This will
significantly prevent one to extract useful information about the neutron-skin thickness
from photon production. How to cancel out the influence of these uncertainties on photon
production is the main task I shall discuss in the following.

To reduce these uncertainties, I propose to use the ratio of photon production from two
reactions to probe the neutron-skin thickness, its definition is
\begin{equation}\label{ratio}
R_{\rm p+^{208}Pb/p+^{40}Ca}(\gamma)\equiv \frac{N_{\gamma}(\rm p+^{208}Pb)}{N_{\gamma}(\rm p+^{40}Ca)}.
\end{equation}
In above equation, the p+$^{40}$Ca reaction with the centrality of 0-100\% is used as
a referential reaction to cancel out the uncertainties from both nucleon-nucleon cross
section and/or photon production probability. This is because the photon production is
mainly determined by proton-neutron colliding number; the proton-neutron colliding
inside the $^{208}$Pb target can be cancelled out by the proton-neutron colliding inside
the $^{40}$Ca, it is naturally that the difference of the photon production from the
incident proton interacting with neutrons inside the skin of $^{208}$ Pb can be shown to
maximal extent. In fact, ratio from two reactions which is usually used in experiments
searching for minute but interesting effects, can reduce maximally not only the systematic
errors but also some 'unwanted' effects \cite{Yong08,Yong11}. Shown in Figs. \ref{med2} and
\ref{prob2} are the ratio of the photon multiplicity from two reactions and the
corresponding relative sensitivity on the neutron-skin thickness with different
nucleon-nucleon cross section, and different photon production probability, respectively,
within the neutron-skin thickness of 0.10 and 0.30fm. It can be found that the ratio of
photon multiplicity from two reactions can almost completely cancel out the uncertainties
from nucleon-nucleon cross section and photon production probability, respectively,
but can keep about 13\%-15\% sensitivity on the neutron-skin thickness.

\begin{figure}[h]
\centerline{\includegraphics[scale=0.35]{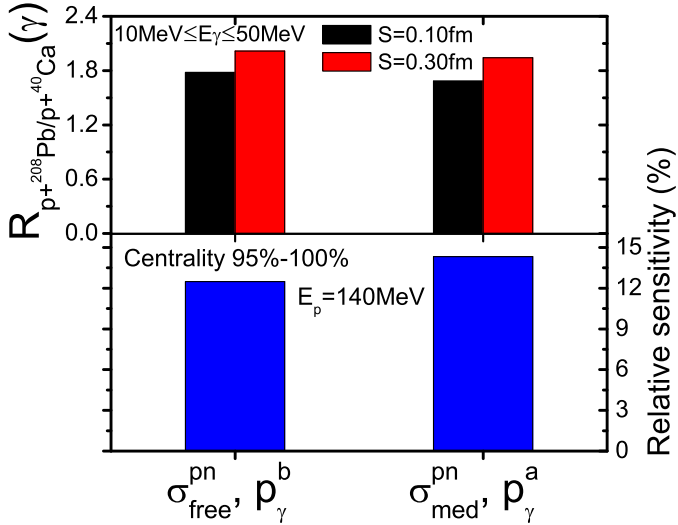}} \caption{(Color
online) The ratio of photon multiplicity from two reactions and the corresponding
relative sensitivity on the neutron-skin thickness with two kinds of setting
in p+Pb reaction, i.e., free-space cross section with photon production probability
formula $p^{b}_{\gamma}$ and in-medium cross section with photon production probability
formula $p^{a}_{\gamma}$ within the neutron-skin thickness of 0.10 and 0.30fm.}
\label{simu}
\end{figure}
\begin{figure}[h]
\centerline{\includegraphics[scale=0.35]{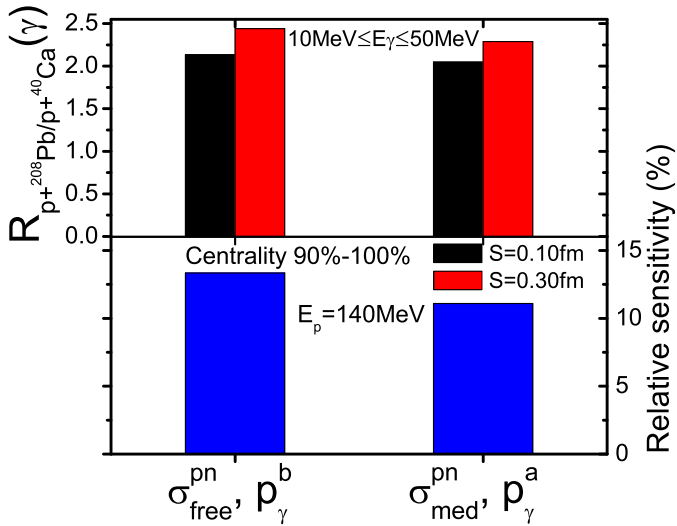}} \caption{(Color
online) Same as Fig. \ref{simu} but with centrality of 90\%-100\%.}
\label{simu2}
\end{figure}

Finally, it is necessary to check whether the ratio of photon multiplicity from two
reactions can simultaneously cancel out these uncertainties from both nucleon-nucleon
cross section and photon production probability since the uncertainties from nucleon-nucleon
cross section and photon production probability exist simultaneously in p+Pb
reaction. Shown in Fig. \ref{simu} are the ratio of photon multiplicity from two
reactions and the corresponding relative sensitivity on the neutron-skin thickness
with two kinds of setting in p+Pb reaction,  i.e., free-space cross section with
photon production probability formula $p^{b}_{\gamma}$ and in-medium nucleon-nucleon
cross section with photon production probability formula $p^{a}_{\gamma}$. It is clear
to see that this ratio can almost completely cancel out these uncertainties from both
nucleon-nucleon cross section and photon production probability simultaneously, but
can preserve about 13\%-15\% sensitivity on the neutron-skin thickness within the
neutron-skin thickness of 0.10 and 0.30fm. On the other hand, considering the experimental
technology limitation of sorting events according to the centrality criteria, the similar
plot with Fig. \ref{simu} but with 90\%-100\% centrality is shown in Fig. \ref{simu2}.
It can be found that this ratio is also sensitivity to the neutron-skin thickness and
shows about 10\%-15\% sensitivities, but is independent approximately of nucleon-nucleon
cross section and photon production probability.

In summary, I have carried out an investigation about the feasibility of probing the
neutron-skin thickness by photon production from neutron-proton bremsstrahlung in
intermediate energy proton-induced reactions. Within the current experimental uncertainty
range of the neutron-skin size of $^{208}$Pb, the p+Pb reaction is performed in different
centralities and at different proton incident energies within a transport model. It is shown
that the energy of about 140MeV for the incident proton and about 95\%-100\% centrality for the
reaction system are the best reaction environment to probe the neutron-skin thickness using
photon production. While the sensitivity of photon production on the neutron-skin
thickness is much smaller than those due to possible uncertainties from both nucleon-nucleon
cross section and photon production probability, the ratio of photon production from two
reactions can almost completely cancel out the influence of these uncertainties simultaneously
but can preserve about 13\%-15\% sensitivity on the neutron-skin thickness. Compared to
other probes involved in nucleon elastic and/or inelastic scattering, photon once produced
can escape almost freely from the strong force environment, it thus can be as a potential
sensitive probe of the neutron-skin thickness in intermediate energy proton-induced reaction.\\

\noindent{\textbf{Acknowledgements}} \\
The author is grateful to Prof. L.W. Chen for stimulating to work on this project, and
Dr. G. C. Yong and Prof. B. A. Li for helpful discussion. This work is supported by the
National Natural Science Foundation of China under grant No.11405128.


\begin{thebibliography}{99}
\bibitem{Fricke95}G. Fricke {\it et al.}, At. Data Nucl. Data Tables \textbf{60}, 177 (1995).
\bibitem{Brown00}B. A. Brown, Phys. Rev. Lett. \textbf{85}, 5296 (2000).
\bibitem{Sun10}X. Y. Sun, D. Q. Fang, Y. G. Ma, X. Z. Cai, J. G. Chen, W. Guo, W. D. Tian, H. W. Wang, Phys. Lett. B \textbf{682}, 396 (2010).
\bibitem{Hoff80}G. W. Hoffmann{\it et al.}, Phys. Rev. C \textbf{21}, 1488 (1980).
\bibitem{Kar02}S. Karataglidis, K. Amos, B. A. Brown, and P. K. Deb, Phys. Rev. C \textbf{65}, 044306 (2002).
\bibitem{Clark03}B. C. Clark, L. J. Kerr, and S. Hama, Phys. Rev. C \textbf{67}, 054605 (2003).
\bibitem{Zen10}J. Zenihiro {\it et al.}, Phys. Rev. C \textbf{82}, 044611 (2010).
\bibitem{Kra99}A. Krasznahorkay {\it et al.}, Phys. Rev. Lett. \textbf{82}, 3216 (1999).
\bibitem{Kra04}A. Krasznahorkay {\it et al.}, Nucl. Phys. A \textbf{731}, 224 (2004).
\bibitem{Kli07}A. Klimkiewicz {\it et al.}, Phys. Rev. C \textbf{76}, 051603(R) (2007).
\bibitem{Car10}A. Carbone, G. Col\`{o}, A. Bracco, L. G. Cao, P. F. Bortignon, F. Camera, and O.Wieland, Phys. Rev. C \textbf{81}, 041301 (2010).
\bibitem{Trz01}A. Trzci\'{n}ska, J. Jastrz\c{e}bski, P. Lubi\'{n}ski, F. J. Hartmann, R. Schmidt, T. von Egidy, and B. K{\l}os, Phys. Rev. Lett. \textbf{87}, 082501 (2001).
\bibitem{Fri03}E. Friedman and A. Gal, Nucl. Phys. A \textbf{724}, 143 (2003).
\bibitem{Jas04}J. Jastrz\c{e}bski, A. Trzci\'{n}ska, P. Lubi\'{n}ski, B. K{\l}os, F. J. Hartmann, T. von Egidy, and S.Wycech, Int. J. Mod. Phys. E \textbf{13}, 343 (2004).
\bibitem{Fri05}E. Friedman, A. Gal, and J. Mare\v{s}, Nucl. Phys. A \textbf{761}, 283 (2005).
\bibitem{Klos07}B. K{\l}os {\it et al.}, Phys. Rev. C \textbf{76}, 014311 (2007).
\bibitem{Fri09}E. Friedman, Hyperfine Interact. \textbf{193}, 33 (2009).
\bibitem{Roca11}X. Roca-Maza, M. Centelles, X. Vi${\rm{\tilde n}}$as, and M. Warda, Phys. Rev. Lett. \textbf{106} 252501 (2011).
\bibitem{Prex1}S. Abrahamyan {\it et al.}, Phys. Rev. Lett. \textbf{108}, 112502 (2012).
\bibitem{Fri12}E. Friedman, Nucl. Phys. A \textbf{896}, 46 (2012).
\bibitem{Fatt13} F. J. Fattoyev and J. Piekarewicz, Phys. Rev. Lett. \textbf{111}, 162501 (2013).
\bibitem{Hor14}C. J. Horowitz, K. S. Kummar, and R. Michaels, Eur. Phys. J. A \textbf{50}, 48 (2014).
\bibitem{Bert88}G. F. Bertsch and S. Das Gupta, Phys. Rep. \textbf{160}, 189 (1988).
\bibitem{Nif90}H. Nifenecker and J. A. Pinston, Annu. Rev. Nucl. Part. Sci. \textbf{40}, 113 (1990).
\bibitem{Cass90}W. Cassing, V. Metag, U. Mosel, and K. Niita, Phys. Rep. \textbf{188}, 363 (1990).
\bibitem{Orte04}R. Ortega, TAPS Collaboration, Nucl. Phys. A \textbf{734}, 541 (2004).
\bibitem{Schu97}Y. Schutz {\it et al.}, TAPS Collaboration, Nucl. Phys. A \textbf{622}, 404 (1997).
\bibitem{Mart99}G. Martinez {\it et al.}, Phys. Lett. B \textbf{461}, 28 (1999).
\bibitem{Ente02}D. d'Enterria {\it et al.}, Phys. Lett. B \textbf{538}, 27 (2002).
\bibitem{Orte06}R. Ortega {\it et al.}, Eur. Phys. J. A \textbf{28}, 161 (2006).
\bibitem{TAPS94}M. Marqu\'{e}s {\it et al.}, Phys. Rev. Lett. \textbf{73} 34 (1994).
\bibitem{Giu06}G. Giuliani and M. Papa, Phys. Rev. C \textbf{73}, 031601 (2006).
\bibitem{Che09}L. W. Chen, B. J. Cai, C. M. Ko, B. A. Li, C. Shen, and J. Xu, Phys. Rev. C \textbf{80}, 014322 (2009).
\bibitem{Che10}L. W. Chen, C. M. Ko, B. A. Li, and J. Xu, Phys. Rev. C \textbf{82}, 024321 (2010).
\bibitem{Wei14}G. F. Wei, B. A. Li, J. Xu, and L. W. Chen, Phys. Rev. C \textbf{90}, 014610 (2014).
\bibitem{Wei15}G. F. Wei, Phys. Rev. C \textbf{91}, 014616 (2015).
\bibitem{IBUU}B. A. Li, C. B. Das, S. Das Gupta, and C. Gale, Phys. Rev. C \textbf{69} (2004) 011603(R); Nucl. Phys. A \textbf{735} (2004) 563.
\bibitem{Bao05}B. A. Li, and L. W. Chen, Phys. Rev. C \textbf{72}, 064611 (2005).
\bibitem{Yong08}G. C. Yong, B. A. Li, and L. W. Chen, Phys. Lett. B \textbf{661}, 82 (2008).
\bibitem{Yong11}G. C. Yong, W. Zuo, and X. C. Zhang, Phys. Lett. B \textbf{705}, 240 (2011).
\bibitem{Gan94}N. Gan {\it et al.}, Phys. Rev. C \textbf{49}, 298 (1994).


\end{thebibliography}
\end{document}